# Evaporative CO$_2$ cooling using microchannels etched in silicon for the future LHCb vertex detector

A. Nomerotski[a,b*], J. Buytart[a], P. Collins[a], R. Dumps[a], E. Greening[b], M. John[b], A. Mapelli[a], A. Leflat[c], Y. Li[d], G. Romagnoli[e], B. Verlaat[a,g]

[a] CERN, CH-1211 Genève 23, Switzerland
[b] University of Oxford, Particle Physics, Denys Wilkinson Building, Keble Road, Oxford OX1 3RH, UK
[c] Skobelitsyn Institute of Nuclear Physics, Moscow State University, Moscow 119991, Russian Federation
[d] Department of Engineering Physics, Tsinghua University, Beijing 100084, China
[e] Department of Mechanical Engineering, University of Genoa, Via all'Opera Pia 15, Genoa 16145, Italy
[g] NIKHEF, Science Park 105, 1098 XG Amsterdam, Netherlands

**Abstract**
The extreme radiation dose received by vertex detectors at the Large Hadron Collider dictates stringent requirements on their cooling systems. To be robust against radiation damage, sensors should be maintained below -20$^o$C and at the same time, the considerable heat load generated in the readout chips and the sensors must be removed. Evaporative CO$_2$ cooling using microchannels etched in a silicon plane in thermal contact with the readout chips is an attractive option. In this paper, we present the first results of microchannel prototypes with circulating, two-phase CO$_2$ and compare them to simulations. We also discuss a practical design of upgraded VELO detector for the LHCb experiment employing this approach.

# Introduction

The LHCb vertex detector (VELO) has pioneered the use of evaporative CO$_2$ cooling for high energy physics since its installation in 2007 [1]. The VELO has 88 silicon sensors, each with 16 readout chips dissipating up to 1 kW in vacuum. The silicon sensors are positioned just 7 mm from the colliding beams and must be kept cool to limit the effect of significant radiation doses. For this reason the cooling system must be not only radiation resistant but low in mass as the cooling components of the modules are within the LHCb acceptance. Also, the unique environment, within the LHC vacuum, requires a perfectly leak tight system.

In 2018 the VELO will be upgraded along with the rest of the LHCb experiment, to operate with increased luminosity, higher integrated radiation dose, and new readout ASICs with higher total heat dissipation [2]. The cooling system must be upgraded, and we present in this paper a solution, which aims to combine recent advances in microchannel cooling technology with the experience already gained with CO$_2$ evaporative cooling.

Two phase microchannel cooling was first proposed with the idea of providing an efficient technique for dissipation of high heat fluxes from microprocessor chips [3]. A system of miniature channels, together with an inlet and outlet manifold, are etched directly into a silicon plate, and are sealed by bonding a second silicon wafer on top. This forms a low mass cooling plate with integrated cooling and high thermal efficiency. Large temperature gradients across the cooling plate are avoided, and a mismatch of the coefficient of thermal expansion between the heat source (silicon sensors and ASICs) and heat sink, is suppressed. This integrated approach is very attractive for high energy physics and is being pursued in a number of domains [4][5]. This paper describes the LHCb approach, which is to integrate the

---

microchannel cooling together with two-phase $CO_2$ as the coolant, thus providing a convenient and radiation hard method to cool the silicon and ASICs to below -20$^o$C over the whole module surface.

Similarly to the current VELO detector, the upgraded design will have at least 21 identical stations positioned along a 1m long beam collision region. Figure 1 shows a possible layout of one station with a cross section for one of the modules in the diagram on the left side. The cross section illustrates the most important point of the approach with the microchannels running through the substrate and $CO_2$ evaporating just under the readout chips, which are the main heat source.

The full station consists of two identical modules, each with four silicon sensors. Each sensor is read out by three Velopix ASICs. All components are mounted on a substrate, which serves as a conduit to the outside world for all electrical and cooling connections. The two sensors and associated readout chips are mounted from the opposite sides of the module as shown in the cross section. The cooling scheme is illustrated for the particular case of microchannel cooling described in detail in the following sections.

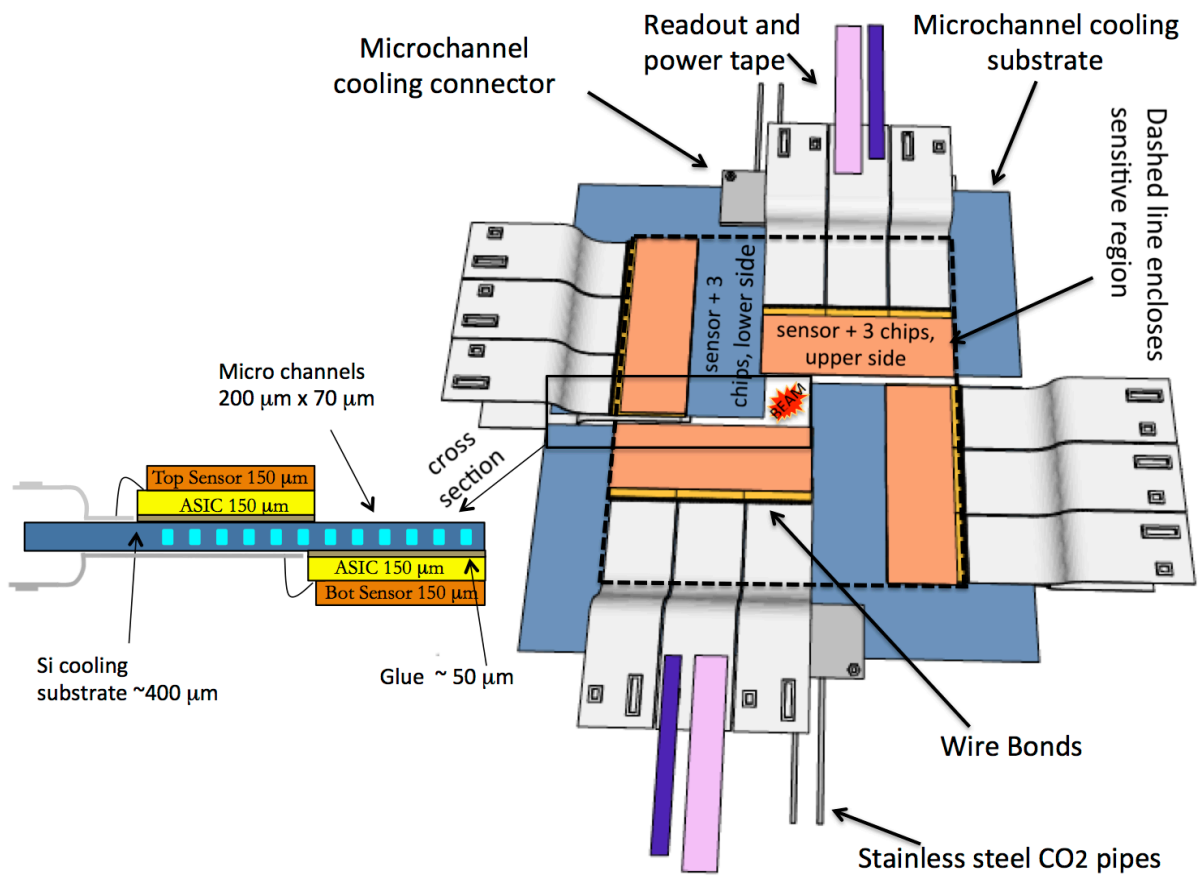

**Figure 1. Possible layout of one upgraded VELO station. The left diagram is the cross section which shows how the microchannel substrate may provide cooling to all ASICs.**

# Module prototype with microchannel cooling

The prototype LHCb samples discussed below were prepared at Ecole Polytechnique Fédérale de Lausanne (Switzerland) employing anodic bonding [6] of a 0.38 mm thick silicon and 0.5 mm thick Pyrex glass wafers. The anodic bonding is performed by applying voltage across the wafers, which are polished and attached to each other. It is a relatively simple procedure with fast turnaround compared to the proper silicon-to-silicon bonding, which is discussed later in the text, but it cannot hold high enough pressure due to the amorphous nature of glass.

The sample has a rectangular shape, 40 mm by 60 mm. Figure 2 shows a drawing and a photograph of the sample. The photograph is taken from the glass side such that the microchannel network is clearly visible. The design, dubbed a "snake" design, has channels which serpentine along the sample allowing the inlet and outlet to be positioned in the same corner of the sample while roughly equalizing the length of all channels. All etching is performed on the silicon wafer only. The microchannels are 70 µm deep while the inlet and outlet are 2 mm diameter through holes separated by 10 mm. Nanoport connectors made of PEEK plastic [7] are glued to the outside surface of the silicon wafer for connection to the cooling system.

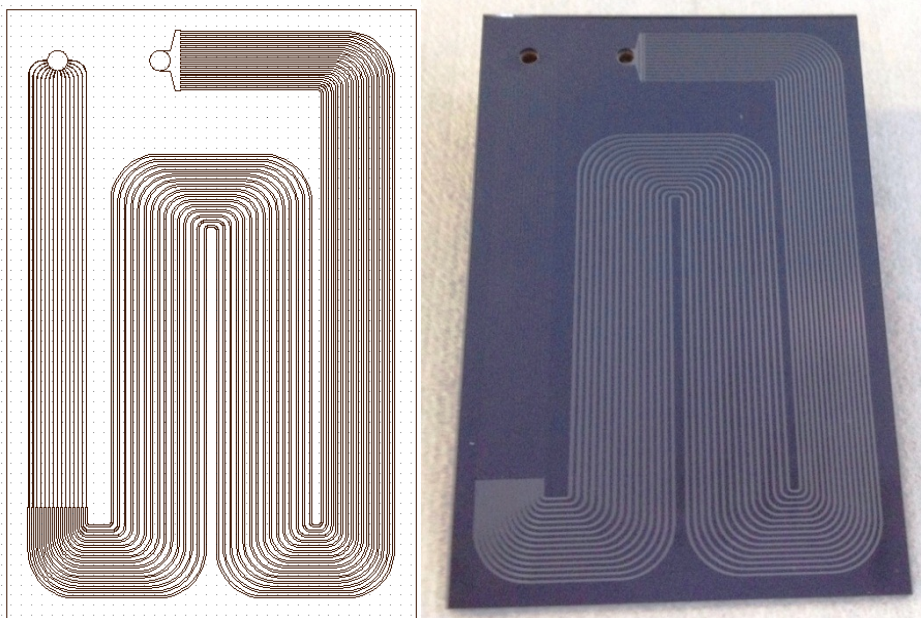

**Figure 2. Drawing and photograph of the "snake" design etched into the silicon substrate.**

In normal operation the modules must be cooled well below zero degrees, however it is mandatory that the system can tolerate a warm up to room temperature, or can even be operated at room temperature for certain test or emergency situations. In this condition the cooling substrate must sustain an overpressure in excess of 65 bar. Relatively large areas of inlet and outlet will result in considerable forces when pressurized, making the manifolds the weakest points to withstand the high pressure. The manifold area needs to be minimized to reduce the force and improve the reliability. The flexibility of the silicon etching process allows this to be achieved by routing the channels directly to the inlet and outlet openings so avoiding conventional, large area manifold areas to distribute and collect the flow from all channels. The design of the inlet and outlet in the "snake" sample is shown in photographs in Figure 3.

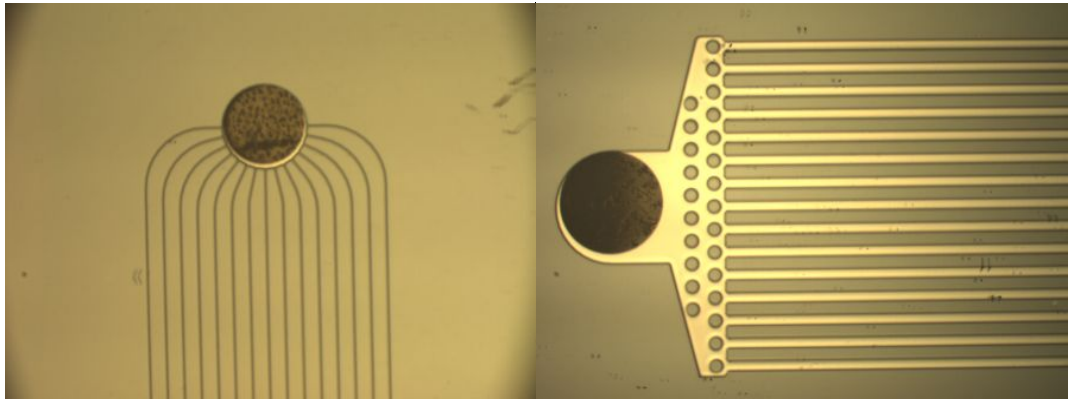
**Figure 3. Photographs of inlet (left) and outlet (right).**

Fifteen microchannels start from the inlet as small capillaries with a width of 30 μm. These inlet capillaries dominate the channel flow impedance and, therefore, ensure a uniform flow distribution across all channels [8]. Without this impedance, power fluctuations in a parallel, two phase system will lead to varying flow rates causing dry-out in the channels with the highest heat load. After 43 mm, the width increases in the evaporator area to 200 μm and remains constant over the full length of the channel until the outlet. The minimum and maximum channel lengths are respectively 195 and 229 mm. Small pillars in the outlet manifold reduce the area and hence the force, and improve the bonding strength. Figure 4 shows SEM photographs of the transition region between the 30 micron wide capillaries and 200 micron wide evaporation channels. The depth of the channels is about 70 micron.

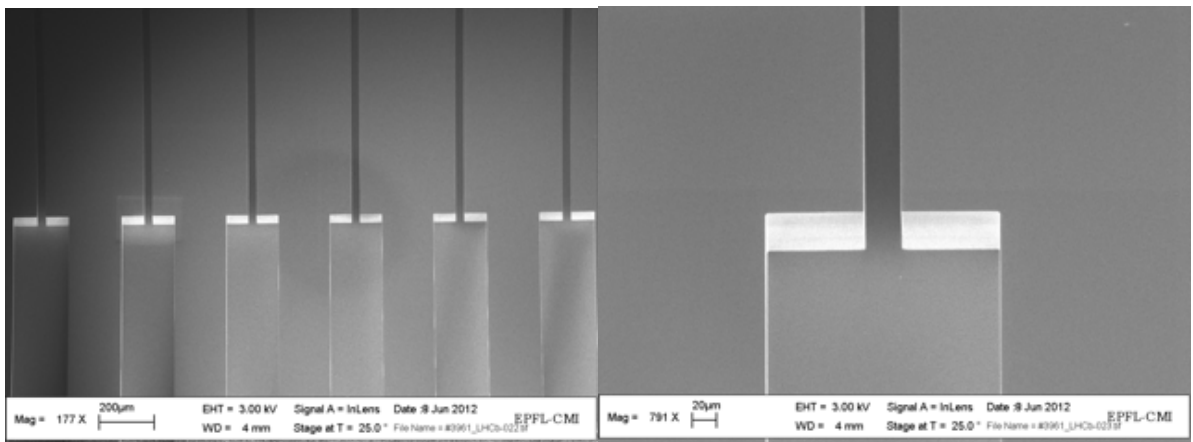
**Figure 4. SEM photographs of the transition region between the 30 micron wide capillaries and 200 micron wide evaporation channels.**

## Experimental setup

The measurements were performed at CERN employing a TRACI $CO_2$ cooling plant [9], specialized vacuum chamber and several microchannel samples with monitored pressure, temperature and flow. The TRACI is capable of circulating liquid $CO_2$ with temperatures down to -40°C whilst maintaining a pressure difference of up to 4 bar between the inlet and outlet. The TRACI was connected to the vacuum chamber with a 3 m long coaxial pipe where the input $CO_2$ flow was surrounded by the output flow for better thermal isolation. The coaxial pipe was interfaced to the vacuum chamber with additional piping employing several valves to allow flexible operation and $CO_2$ refill. The inlet and outlet pressure was monitored before entering the vacuum chamber. The vacuum chamber, which maintains a vacuum pressure of about 1 mbar, is needed to avoid ice formation at low temperature. The required electrical connections were implemented with vacuum-tight feed-throughs. Monitoring was carried out with modules from National Instruments controlled using LabView software.

Figure 5 shows a photograph of assembled microchannel sample from the connector side with connectors, a heater and temperature sensors. The connector area was strengthened with a 2 mm thick steel plate with two connector holes and two mounting holes. The plate was glued to the sample on the top of the connectors to restrain them and protects the silicon structure from possible torque stresses caused by the connection to the steel piping, which is quite stiff. A $(15 \times 7 \times 0.5)$ mm$^3$ piece of silicon, not visible in the photograph, was glued above the inlet and outlet holes on the other side to further strengthen this area.

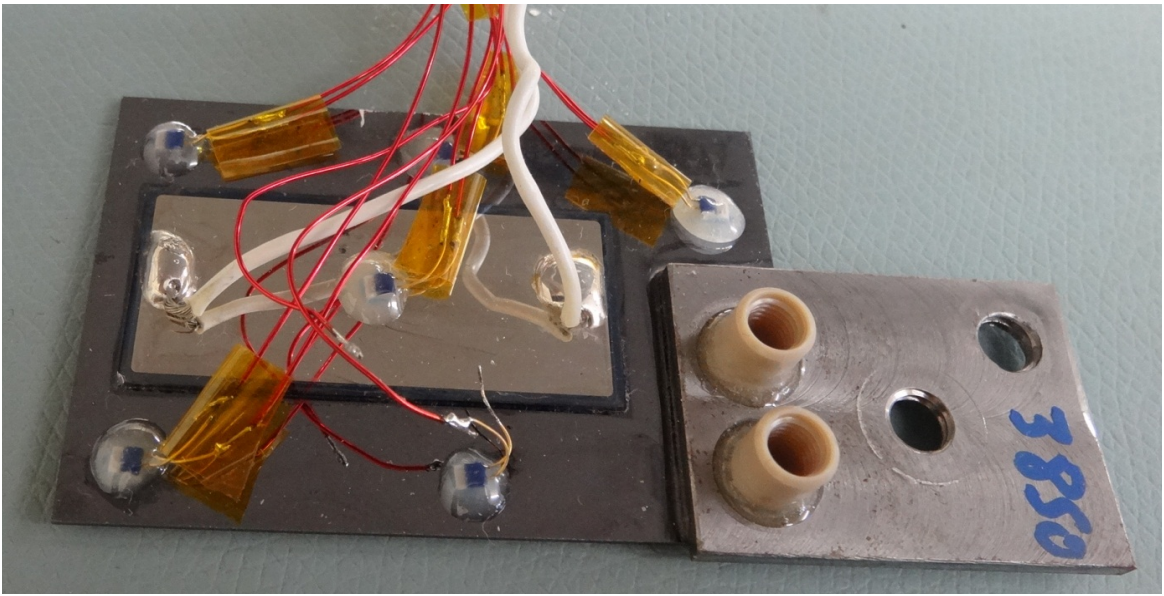

**Figure 5. The heater, temperature sensors, Nanoport connectors and a steel handle glued to the substrate, viewed from the silicon side.**

The sample was equipped with a heater made of metallized silicon and several temperature sensors, shown in the photograph. The 6.2 Ohm $(40 \times 17)$mm$^2$ heater was located in the centre of the sample above two middle strands of microchannels, see Figure 2. The temperature sensors were based on platinum 100 Ohm resistors read out using a four-point connection. Six temperature sensors were glued along the channel length (one on the top of the heater) and two more sensors on the inlet and outlet pipes about 10 mm away from the Nanoport connectors.

## Test results

The sample was connected to TRACI and tested in a variety of conditions, which are described in detail below.

For $CO_2$ cooling operation a microchannel needs to withstand the high pressure of $CO_2$ at start-up. This is at least 65 bar as this is the saturation pressure of $CO_2$ at room temperature, 25$^o$C. In a normal start-up the system is pressurized and warm liquid will flow through the micro channels. For the initial test samples it was chosen to start-up under low pressure conditions not to risk a loss of a sample due to overpressure. To establish the $CO_2$ circulation under low pressure the sample and piping first needs to be cooled down with a gas flow other than the usual liquid flow. The microchannels represent considerable impedance to the gas flow making the cooldown process with gas quite slow. To speed up the cooldown a bypass route with an adjustable valve was installed in parallel to the microchannel substrate in the vacuum chamber. In the beginning the valve was fully open allowing for an increased $CO_2$ flow to cool down the transfer piping from TRACI to the experimental setup. After this the valve was partially closed to start the $CO_2$ circulation through the microchannel substrate. This corresponds to the upward temperature spike at 100 seconds in Figure 6, which shows the cooldown process. The sudden drop in the inlet temperature at 260 seconds corresponds to the moment when liquid $CO_2$ arrives to the substrate. After this the substrate temperatures drift down until settled at $0^{o}C$ when the circulation is established. The $CO_2$ temperature is in direct relation to the $CO_2$ pressure, in this particular case 35 bar. The adjustable valve was also used during the measurements to regulate the pressure drop across the sample hence regulating the $CO_2$ flow through the sample.

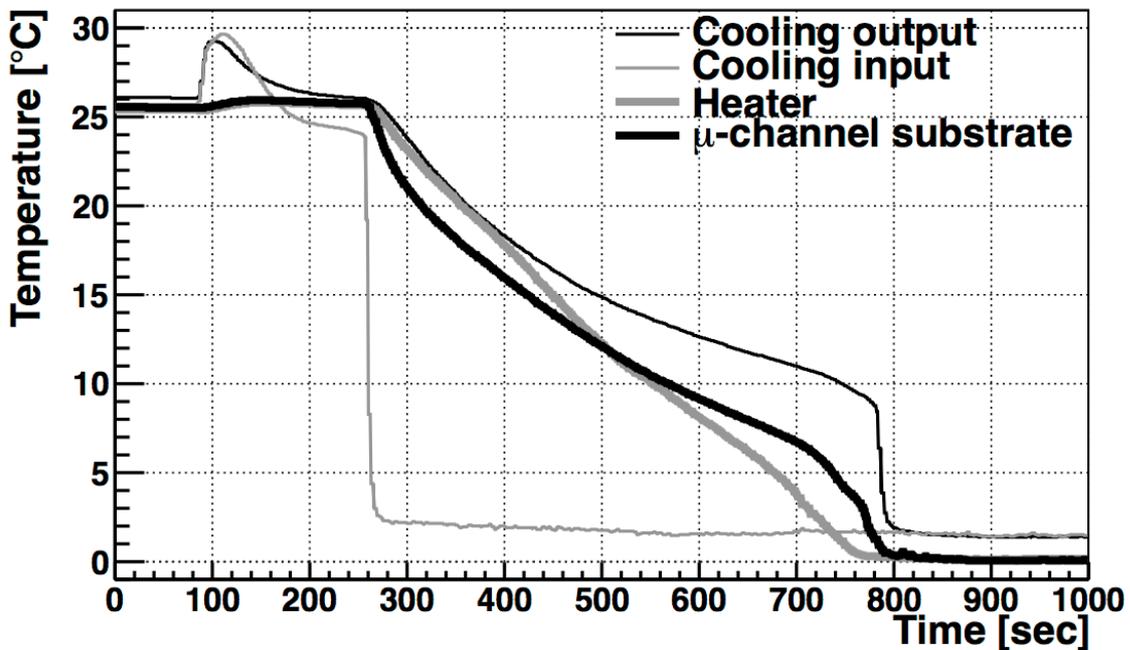

**Figure 6. The cooldown time sequence for the microchannel substrate.**

To determine the maximum power, which can be dissipated by the cooled sample, the voltage across the heater was ramped up until the dry-out condition was achieved. A typical time sequence of the monitored temperatures is shown in Figure 7. The four time-traces in the figure are the inlet, outlet, heater and substrate temperatures. The substrate temperature was monitored near the outlet after the heater. The histogram in the top graph corresponds to the power dissipated in the heater. The inlet temperature is somewhat higher than the other temperatures due to environmental heat leak to the incoming liquid. It is limited by the local saturation temperature, which is higher because of the pressure drop across the microchannels. During the dry-out all $CO_2$ liquid in microchannels is evaporated and the outlet temperature starts to rise, as can be seen in the figure after 430 seconds. It should be noted that the temperature of the heater remains constant for long time after the dry-out started. The maximum power achieved before the dry-out was equal to 12.9 W. These measurements were performed at the outlet pressure of 35 bar and total $CO_2$ mass flow through the microchannels of 0.15 g/sec.

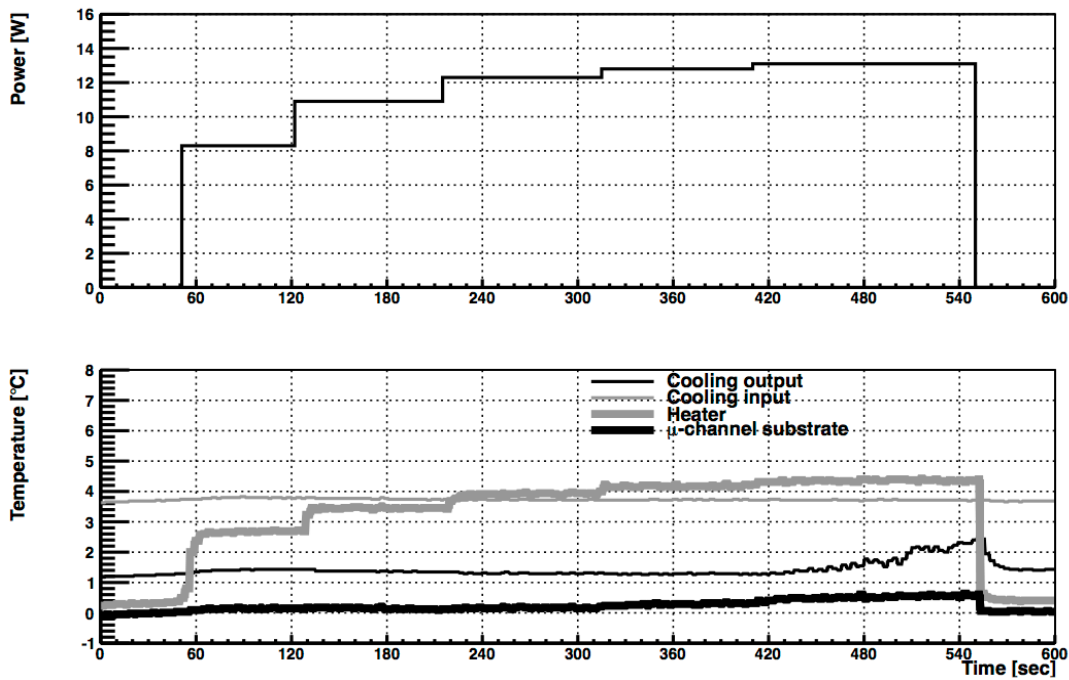

Figure 7. Top: the power provided by the variable heater. Bottom: temperature trends on the cooled substrate. The four trends correspond to four sensors placed along the path of the $CO_2$.

The same data can be presented as difference between the heater and outlet temperatures versus the power dissipated by the heater, see Figure 8. The maximum difference is equal to 2.9°C at 12.9 W. As expected the dependence is linear until the onset of the dry-out, where a small non-linearity is observed, explained by transition to a different cooling regime.

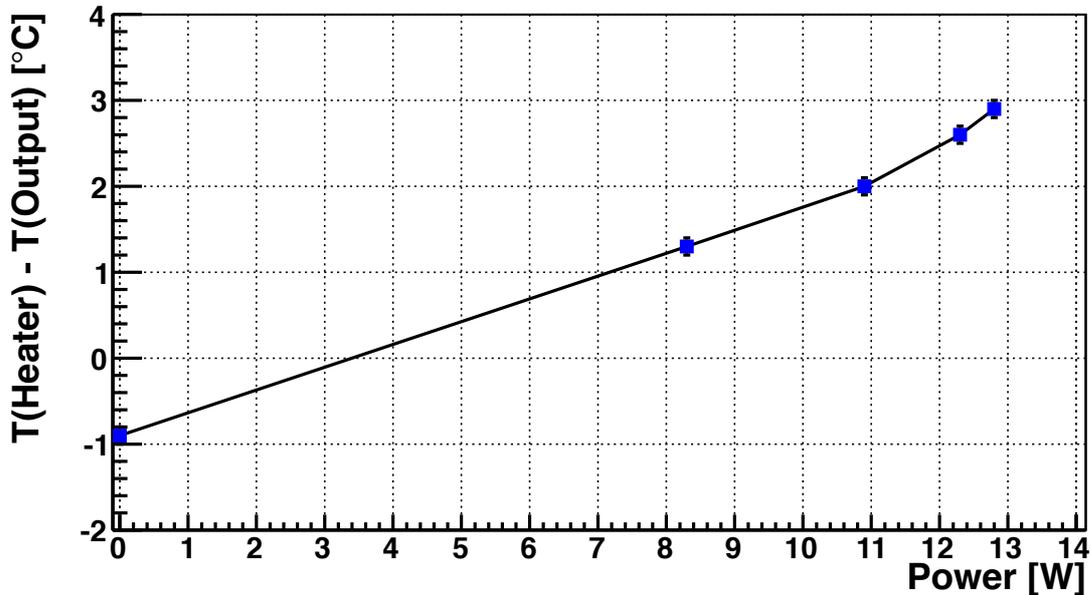

Figure 8. Temperature difference between the heater and cooling output versus dissipated power.

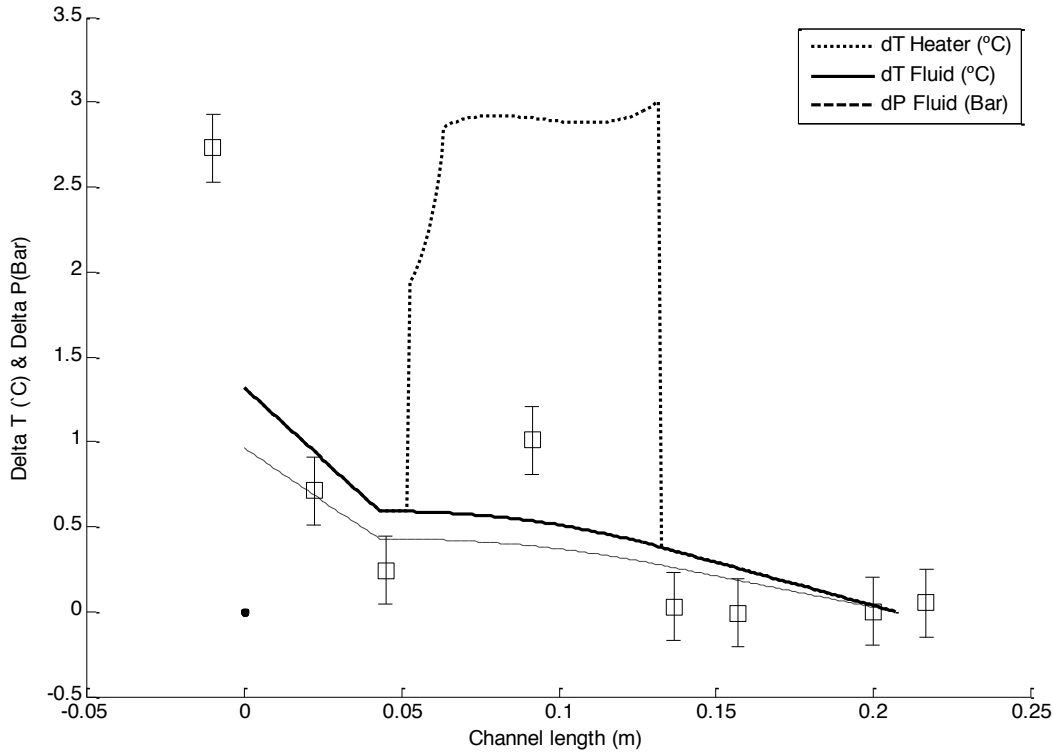

**Figure 9. A comparison of simulation of the temperature difference with respect to the outlet temperature along the channel length (solid, dotted and dashed lines) and the data (squares with error bars), see discussion in text.**

The distribution of temperature along the channel length, shown in Figure 9, is another interesting observable. These measurements were performed at -12º C, the power dissipated in the heater was 2.9 W and the total $CO_2$ mass flow was 0.12 g/sec. The inlet temperature, measured on the steel pipe about 10 mm before the Nanoport connector, is the highest due to higher pressure at this point and environmental warm-up of the $CO_2$ liquid. The results are compared to the simulations performed using the CoBra software [8]. The simulation algorithm predicts the flow regime in each point of the channel and employs analytical calculations for different regimes allowing determination of pressure and temperatures along the channel. The channel is split into sections along the length corresponding to different widths and to the presence of the heater. The following section assignment is used: 30 μm capillaries; 200 μm evaporation channel before the heater; heater; 200 μm evaporation channel after the heater. The pressure drop along the channel length is faster in the initial section with small capillaries slowing down in the 200 μm wide channels. The temperature does not change much until the section with the heater where the $CO_2$ boils. In this section the temperature of the channel wall is higher than that of the core liquid. The systematic variations, especially obvious in the heater section, can be explained by the limited applicability of the model to microchannels, which have rather small dimensions, and by the heat spread across the whole substrate due to the thermal conductivity of silicon.

This cooling substrate was also used to cool a Medipix3 readout chip [10] bump-bonded to a silicon sensor. The dimensions of the chip are 17.3 x 14.1 mm$^2$. It was glued on to a kapton hybrid with total thickness of 0.1 mm, which provided all electrical connections. The hybrid was mounted on a 0.5 mm thick, (4×4) cm$^2$ silicon plate, which was attached to the microchannel cooling substrate with a heat transfer paste (MS1001, Schaffner). The temperature was monitored on the inlet and outlet stainless steel pipes, 10 mm from the Nanoport connectors, on the top of the silicon sensor and on the kapton hybrid. The assembly was cooled to -27°C using the same cooling plant as described above. The inlet pressure was equal to 17 bar. The pressure drop across the microchannels was 3.1 bar corresponding to the total mass flow of 0.18 g/sec. The Medipix3 chip was powered and read out for several minutes providing a 1.5 W heat load. Figure 10 shows the time dependence of the four temperatures with increase of the silicon and hybrid temperatures corresponding to the operating Medipix3. The temperature difference between the ON and OFF periods was 3.2 and 0.4°C, for the top of the silicon sensor and for the kapton hybrid respectively.

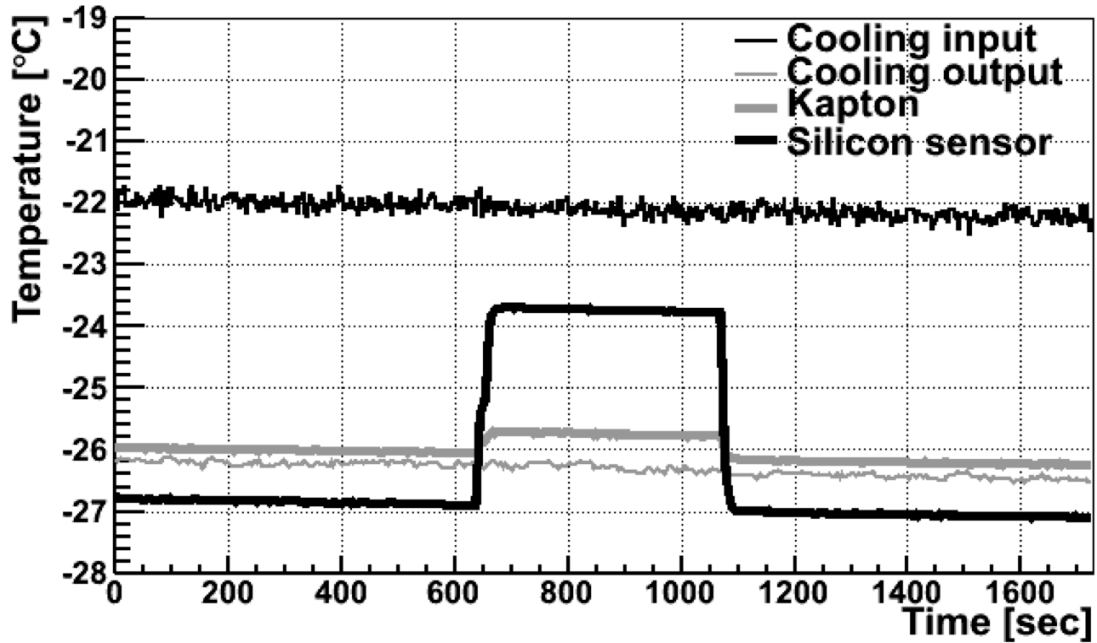

**Figure 10.** Time dependence of four temperatures: inlet, outlet, top of the silicon sensor and kapton hybrid. The increase of the temperature is due to the heat load provided by the operating Medipix3 chip.

## Discussion

The above result is the proof-of-principle demonstration that the evaporative $CO_2$ cooling can be implemented in the microchannels etched in silicon. The small temperature difference achieved between the heat source and coolant is unprecedented for similar systems and will allow for very efficient cooling of silicon detectors with minimal material budget.

One of the most important considerations in cooling of silicon damaged by radiation is the thermal runaway [11]. The silicon leakage current has a strong dependence on the temperature and, after a large irradiation dose, could dissipate power comparable to the power dissipated by the readout electronics. This creates conditions for a positive feedback mechanism with uncontrollable temperature rise. Maintaining a small temperature difference between the silicon sensor and coolant is critical to mitigate this effect. A large temperature difference would impose stringent requirements on the cooling system design and would make the thermal behaviour very sensitive to small variations in its operation conditions.

The Thermal Figure of Merit (TFoM = Area×($T_{heater}$-$T_{out}$)/Power) for the microchannel samples presented here is equal to 1.5, 2.6 and 4.2 $cm^2K/W$, derived respectively from the results in Figures 7, 9 and 10. The TFoM variation in the three cases is explained by different measurement conditions and different geometry. For example, in the last measurement the temperature was monitored on the top of the silicon detector, separated from the cooling substrate by a considerable amount of material including the Medipix3 chip, while in the first two cases the temperature was monitored directly on the heater surface, which was glued directly to the cooling substrate. These results can be favourably compared, for example, to the ATLAS Inner B-Layer design [12], which has TFoM = 13 cm$^2$K/W. The difference is explained by the smaller ΔT achieved by the microchannel design.

## Future plans

Further work is proceeding in two directions: the detailed investigation of the cooling performance and its comparison with simulations; and the development of practical designs for the upgraded VELO detector. The latter will require samples employing the Si-Si or Si-SiO$_2$-Si fusion bonding [13], which can withstand the pressure in excess of 100 bar. In this case the cooling system will be able to warm to the room temperature with some safety margin. An important step in this direction was demonstration that a "snake" cooling substrate with 0.38 mm Si and 2.0 mm glass bonded using anodic process can withstand the pressure of up to 67 bar. An R&D programme is being pursued with several vendors that are able to manufacture samples with fusion bonding with the total silicon thickness of 0.4 mm.

Connectivity is another issue that must be solved since the Nanoport connectors used for these proof-of-principle measurements are not rated for high pressure, nor for vacuum applications. Figure 11 illustrates a possible connector for the VELO module. A metal connector has two parts, attached by three screws, and clamp the cooling substrate in between. The holes in the connector are aligned with the inlet and outlet holes of the module. Two O-rings ensure leak-tight joints between the microchannel network and the connectors. Two metal pipes penetrate into the holes to provide connectivity to the outside world. Though not apparent in this concept drawing, all possible will be done to minimize the connector mass for the final design.

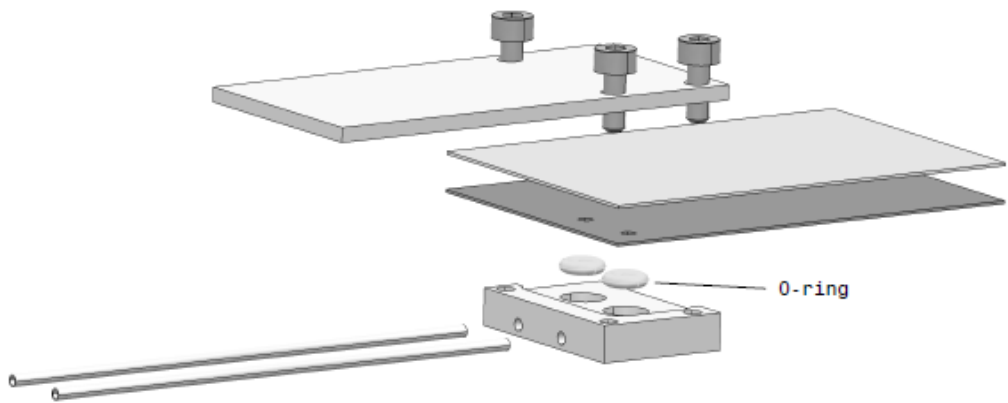

**Figure 11. Concept drawing of the new connector.**

A possible layout of microchannels for the upgraded VELO module is shown in Figure 12. In essence, the layout is a combination of two samples described previously into a single object. The inlet accommodates 30 microchannels, 30 μm wide; 15 of them are going to one side and the remaining 15 are going to the other side of the module. After 75 mm, and before the channels turn under the readout chips the width increases to 200 μm as in the original design. The channels end with two outlets. The inlet and two outlets will be attached to a connector similar to the one described above. The two outlets will be connected to each other inside the connector.

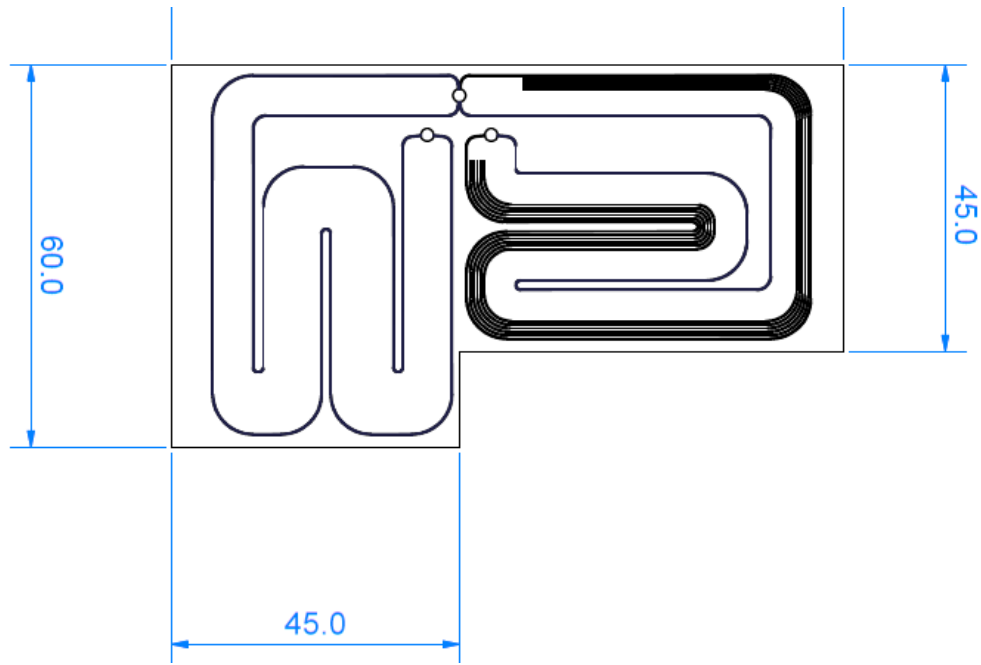

**Figure 12. Concept drawing of the upgraded VELO module with microchannel cooling.**

## Conclusions

We reported on the pioneering, proof-of-principle measurements employing evaporative $CO_2$ cooling in silicon microchannels in the context of the upgraded LHCb VELO detector. We were able to remove 12.9 W of power before dry-out from a prototype sample, which models a quarter of one VELO station. This meets stringent cooling specifications for the upgraded detector providing minimal temperature gradients of less than 4$^o$C in the module. The measurements were performed at a range of parameters with maximum operational pressure of 67 bar at room temperature and minimal temperature of -27$^o$C corresponding to the $CO_2$ pressure of 17 bar.

We also described further steps and a practical implementation of this approach for the VELO upgrade. A new connector, capable of withstanding the high pressure required for operation at room temperature is illustrated, as well as a possible layout of microchannels in a module.

## Acknowledgements

We are thankful to P. Lau for preparing the "snake" design drawings; to J. Noel, P. Petagna and T. Singh for discussions and preparation of the samples used in these measurements; R.N. Collins for help with preparing the diagrams and to N. Nakatsuka for help with measurements.

## References

[1] Verlaat B. et al,"$CO_2$ Cooling for the LHCb-VELO Experiment at CERN", 8th IIF/IIR Gustav Lorentzen Conference on Natural Working Fluids, Copenhagen, Denmark , 2008.

[2] The LHCb collaboration. "Letter of Intent for the LHCb Upgrade", CERN-LHCC-2011-001, 2011.

[3] J.R. Thome, G. Ribatski, "State-of-the art of flow boiling and two-phase flow of $CO_2$ in macro- and micro-channels", Int. J. Refrig. 28 (2006), pp. 1149–1168.

[4] Kandlikar S. G., "High Flux Heat Removal with Microchannels—A Roadmap of Challenges and Opportunities", Heat Transfer Engineering, 26(8), pp. 5–14, 2005.

[5] Mapelli A. et al., "Micro-channel cooling for high-energy physics particle detectors and electronics", Thermal and Thermomechanical Phenomena in Electronic Systems (ITherm), 13th IEEE Intersociety Conference, pp. 677-683, 2012.

[6] Wallis G. and Pomerantz D. I., "Field assisted glass-metal sealing", J. Appl. Phys., 40 (1969) 3946-49.

[7] Upchurch Scientific; www.upchurch.com.

[8] Verlaat B., Noite J., "Design considerations of long length evaporative $CO_2$ cooling lines", 10th IIF/IIR Gustav Lorentzen Conference on Natural Working Fluids, Delft, The Netherlands 2012.

[9] Verlaat B. et al, "TRACI, a multipurpose $CO_2$ cooling system for R&D", 10th IIF/IIR Gustav Lorentzen Conference on Natural Working Fluids, Delft, The Netherlands 2012.

[10] Ballabriga, R., Campbell, M., Heijne, E., Llopart, X., Tlustos, L., Wong, W., "Medipix3: A 64k pixel detector readout chip working in single photon counting mode with improved spectrometric performance", Nuclear Inst. and Methods in Physics Research A, vol. 633 May, 2011. p. S15-S18.

[11] T. Kohriki et al, "First Observation of Thermal Runaway in the Radiation Damaged Silicon Detector", IEEE Transactions on Nuclear Science, vol. 43, no. 3, June 1996.

[12] ATLAS Insertable B-Layer Technical Design Report, CERN-LHCC-2010-013, 15 September 2010.

[13] Ploessl A. and Krauter G., "Wafer direct bonding, tailoring adhesion between brittle materials", Mat. Sci. Eng. R25 (1999) 1-88.